\documentclass[%
 reprint,
superscriptaddress,
 amsmath,amssymb,
 aps,
prx,longbibliography,
]{revtex4-2}
\usepackage{graphicx}
\usepackage{dcolumn}
\usepackage{bm}
\usepackage{multirow}
\usepackage{float}

\AtBeginDocument{%
	\newwrite\bibnotes
	\def\bibnotesext{SnSe.bib}
	\immediate\openout\bibnotes=\jobname\bibnotesext
	\immediate\write\bibnotes{@CONTROL{REVTEX42Control}}
	\immediate\write\bibnotes{@CONTROL{%
			apsrev42Control,author="08",editor="1",pages="1",title="0",year="1"}}
	\if@filesw
	\immediate\write\@auxout{\string\citation{apsrev42Control}}%
	\fi
}%
\begin{document}
\title{Determination of nonthermal bonding origin of a novel photoexcited lattice instability in SnSe}
\author{Yijing Huang}
\affiliation{Stanford Institute for Materials and Energy Sciences, SLAC National
Accelerator Laboratory, Menlo Park, California 94025, USA}
\affiliation{Department of Applied Physics, Stanford University, Stanford, California
94305, USA}
\affiliation{Stanford PULSE Institute, SLAC National Accelerator
Laboratory, Menlo Park, California 94025, USA}
\email{huangyj@stanford.edu}

\author{Samuel Teitelbaum}%
\affiliation{Stanford Institute for Materials and Energy Sciences, SLAC National
Accelerator Laboratory, Menlo Park, California 94025, USA}
\affiliation{Stanford PULSE Institute, SLAC National Accelerator
Laboratory, Menlo Park, California 94025, USA}
\affiliation{Department of Physics, Arizona State University, Tempe, AZ 85287, USA}

\author{Shan Yang}%
\affiliation{Department of Mechanical Engineering and Materials Science, Duke University, 
	Durham, North Carolina 27708, USA}

\author{Gilberto De la Pe\~na}
\affiliation{Stanford Institute for Materials and Energy Sciences, SLAC National
Accelerator Laboratory, Menlo Park, California 94025, USA}
\affiliation{Stanford PULSE Institute, SLAC National Accelerator
Laboratory, Menlo Park, California 94025, USA}

\author{Takahiro Sato}
\affiliation{Linac Coherent Light Source, SLAC National Accelerator Laboratory, Menlo Park, California
94025, USA}

\author{Matthieu Chollet}
\affiliation{Linac Coherent Light Source, SLAC National Accelerator Laboratory, Menlo Park, California
94025, USA}

\author{Diling Zhu}
\affiliation{Linac Coherent Light Source, SLAC National Accelerator Laboratory, Menlo Park, California 94025, USA}

\author{Jennifer L. Niedziela}
\affiliation{Department of Mechanical Engineering and Materials Science, Duke University, 
Durham, North Carolina 27708, USA}
\affiliation{Materials Science and Technology Division, Oak Ridge National Laboratory, Oak Ridge, Tennessee 37831, USA}

\author{Dipanshu Bansal}
\affiliation{Department of Mechanical Engineering and Materials Science, Duke University, 
Durham, North Carolina 27708, USA}
\affiliation{Department of Mechanical Engineering, Indian Institute of Technology Bombay, Mumbai, MH 400076, India}

\author{Andrew F. May}
\affiliation{Materials Science and Technology Division, Oak Ridge National Laboratory, Oak Ridge, Tennessee 37831, USA}

\author{Aaron M. Lindenberg}
\affiliation{Stanford Institute for Materials and Energy Sciences, SLAC National
Accelerator Laboratory, Menlo Park, California 94025, USA}
\affiliation{Stanford PULSE Institute, SLAC National Accelerator
Laboratory, Menlo Park, California 94025, USA}
\affiliation{Department of Materials Science and Engineering, Stanford University, Stanford, CA 94305, USA}

\author{Olivier Delaire}
\affiliation{Department of Mechanical Engineering and Materials Science, Duke University, 
Durham, North Carolina 27708, USA}
\affiliation{Department of Physics, Duke University, 
Durham, North Carolina 27708, USA }
\affiliation{Department of Chemistry, Duke University, 
Durham, North Carolina 27708, USA }

\author{Mariano Trigo}
\affiliation{Stanford Institute for Materials and Energy Sciences, SLAC National
Accelerator Laboratory, Menlo Park, California 94025, USA}
\affiliation{Stanford PULSE Institute, SLAC National Accelerator
Laboratory, Menlo Park, California 94025, USA}
\author{David A. Reis}
\affiliation{Stanford Institute for Materials and Energy Sciences, SLAC National
Accelerator Laboratory, Menlo Park, California 94025, USA}
\affiliation{Department of Applied Physics, Stanford University, Stanford, California
94305, USA}
\affiliation{Stanford PULSE Institute, SLAC National Accelerator
Laboratory, Menlo Park, California 94025, USA}
\affiliation{Department of Photon Science, Stanford University, Stanford, California
94305, USA}
\email{dreis@stanford.edu}
\begin{abstract}
Interatomic forces that bind materials are largely determined by an often complex interplay between the electronic band-structure and the atomic arrangements to form its equilibrium structure and dynamics. As these forces also determine the phonon dispersion, lattice dynamics measurements are often crucial tools for understanding how materials transform between different structures.  This is the case for the mono-chalcogenides which feature a number of lattice instabilities associated with their network of resonant bonds and a large tunability in their functional properties.  
SnSe hosts a novel lattice instability upon above-bandgap photoexcitation that is distinct from the distortions associated with its high temperature phase transition, demonstrating that photoexcitation can alter the interatomic forces significantly different than thermal excitation.
Here we report decisive time-resolved X-ray scattering-based measurements of the nonequlibrium lattice dynamics in SnSe.   
By fitting interatomic force models to the excited-state dispersion, we determine this instability as being primarily due to changes in the fourth-nearest neighbor bonds that connect bilayers, with relatively little change to the intralayer resonant bonds. In addition to providing critical insight into the nonthermal bonding origin of the instability in SnSe, such measurements will be crucial for understanding and controlling materials properties under non-equilibrium conditions.

\end{abstract}

\maketitle

  \begin{figure*}
	\centering
	\includegraphics{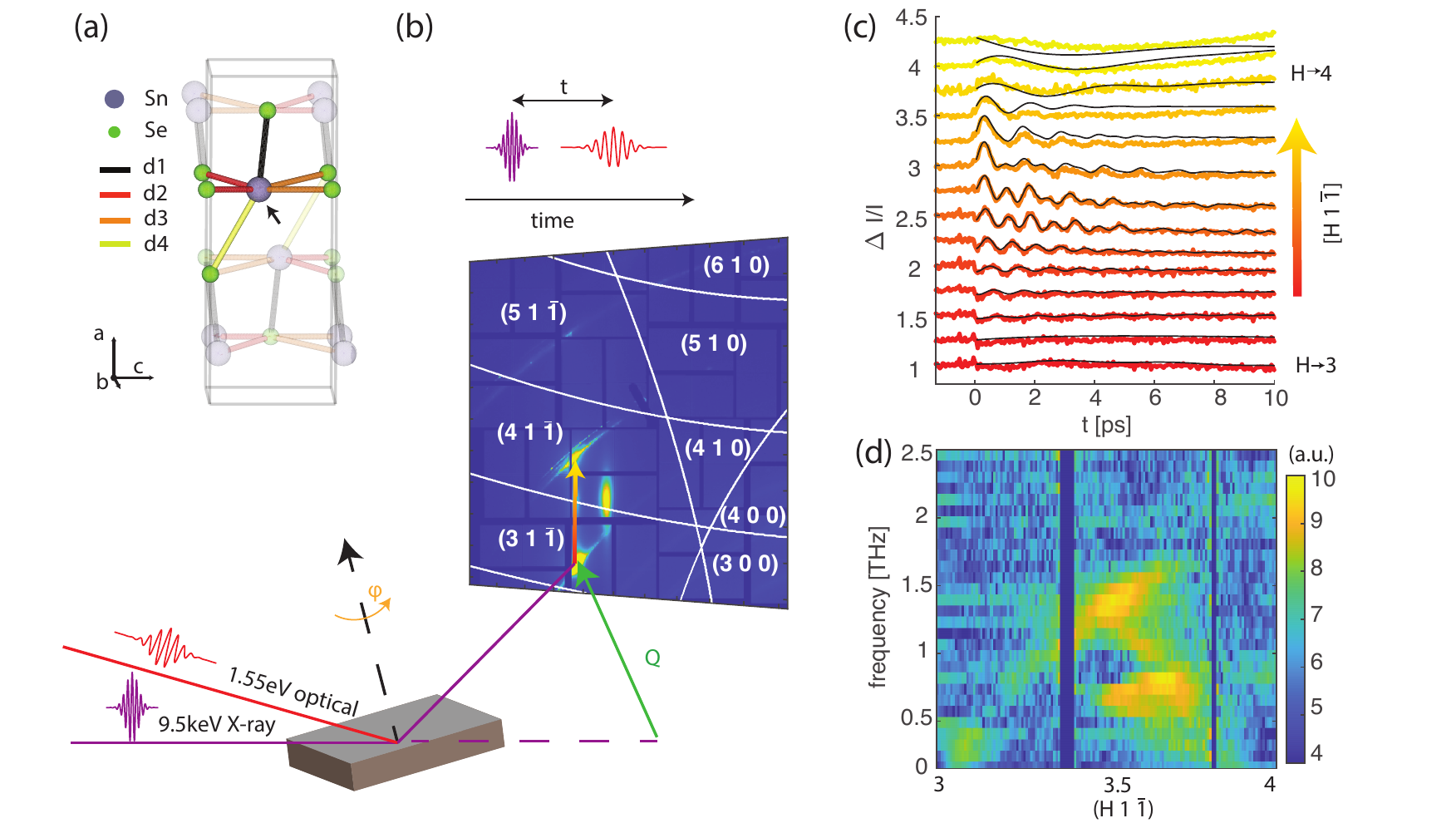}
	\caption{\textbf{Time resolved diffuse scattering of SnSe.}
	\textbf{a} The unit cell of the Pnma phase of SnSe illustrating the $d_1-d_4$ bonds derived from nearest neighbor bonds in the parent cubic structure. The black arrow indicates the space opened up by the $d_4$ bond tilting away from $\mathbf{a}(x)$ direction.
	\textbf{b} The experimental setup. The sample can be rotated around its normal by an azimuthal angle, and a 2D detector captures the diffuse scattering intensity as function of delay $t$ between the optical pump and X-ray probe. The green vector shows a scattering wavevector (momentum transfer) $\textbf{Q}$, associated with scattering on a particular pixel on the detector. The detector image shows the typical intensity pattern for a fixed azimuth $\varphi$, without pump. White lines represent the Brillouin zone boundaries of the Pnma structure.  
	\textbf{c} Time dependence of the relative intensity for representative $\textbf{Q}$ along ($H$1$\bar{1}$), $H\in$[3,4].
	Black lines show reconstructions based on linear prediction.
	\textbf{d} The magnitude of the Fourier transform of time traces as those shown in Fig.1\textbf{c}.
	}  
	\label{fig:general_procedure}
\end{figure*}

\begin{figure*}
	\centering
	\includegraphics{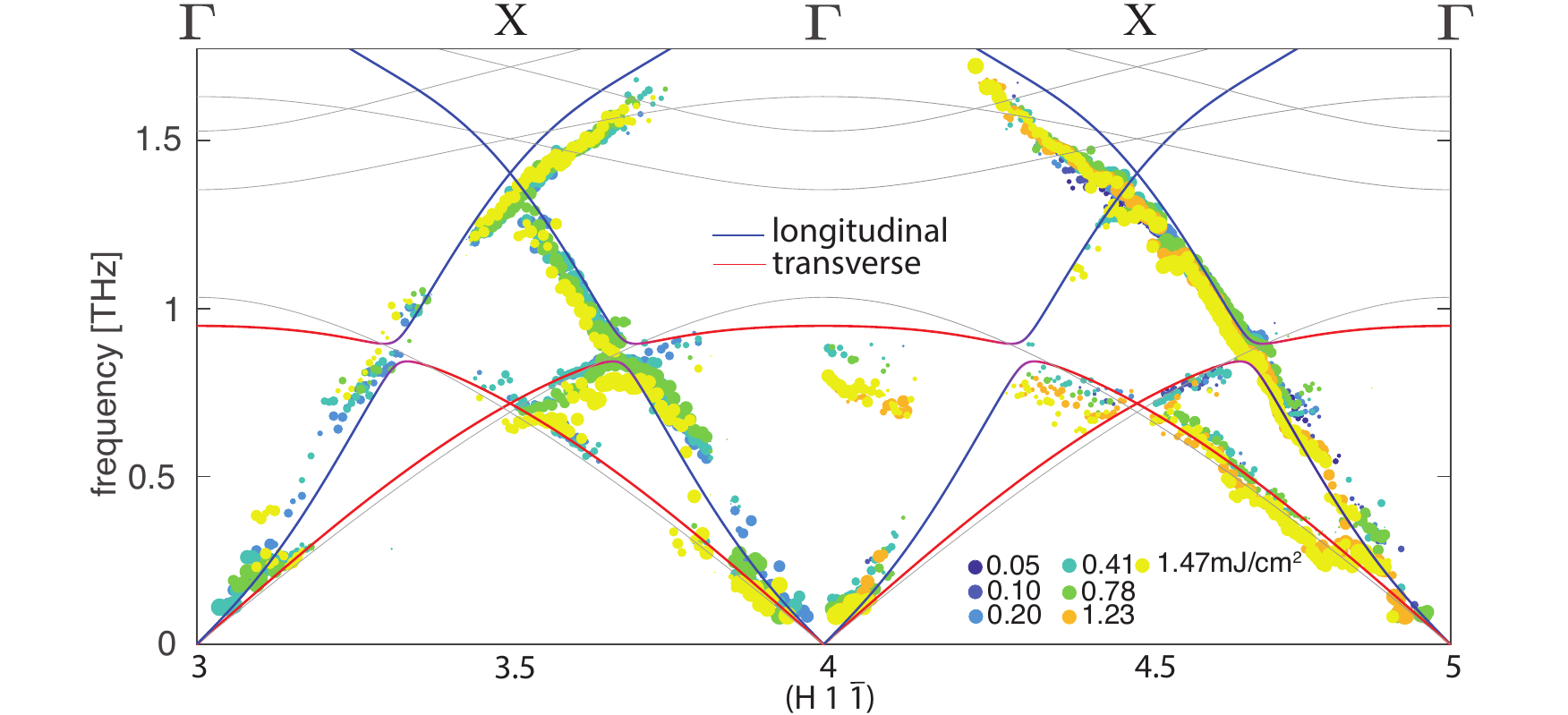}
	\caption{\textbf{Photoexcited phonon dispersion of SnSe.}
	Low frequency region of the dispersion along for phonons propagating along the $a$-direction ($\textbf{Q}$= ($H$1$\bar{1}$), $H\in$[3,5]) extracted from linear prediction (see text). The solid lines are the ground state phonon dispersion based on density-functional theory calculations. Red and blue lines represent the lowest two transverse and longitudinal branches polarized along $\bf{c}$-axis and $\bf{a}$-axis, respectively. 
	}  
	\label{fig:full_dispersion}
\end{figure*}

\begin{figure}
	\centering
	\includegraphics{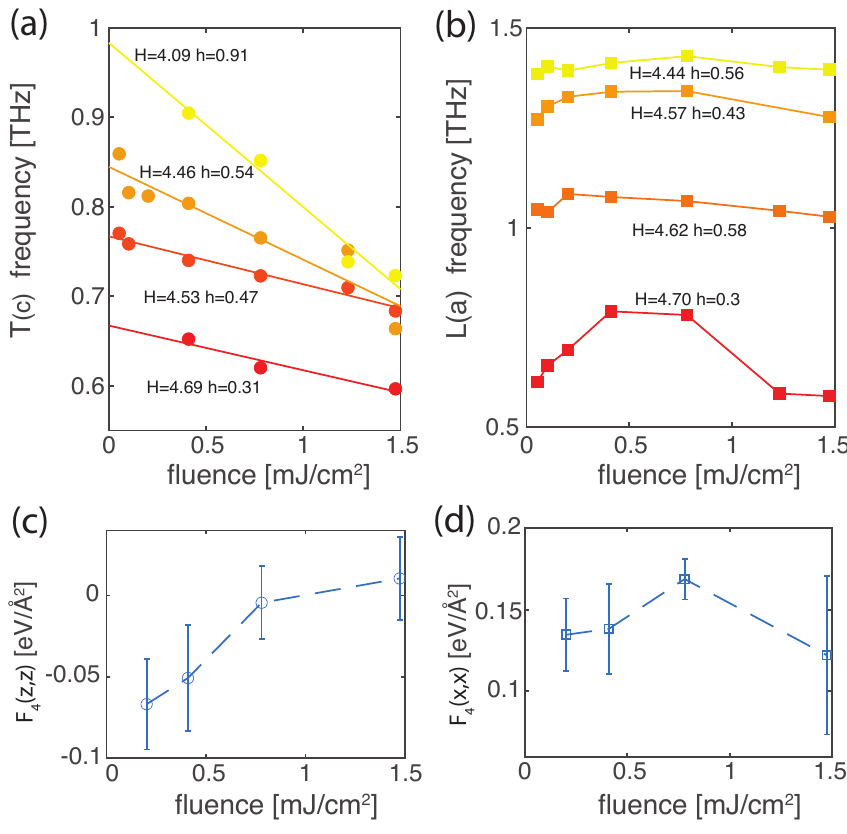}
	\caption{ \textbf{Fluence dependence of photoexcited SnSe phonons and force constants.}
\textbf{a} Nearly linear fluence dependent softening of $T(\mathbf{c})$ phonon along (H1$\bar{1}$), $H\in$[4,5] for a subset of $\mathbf{Q}$ points. 
\textbf{b} Non-monotonic fluence dependence of $L(\mathbf{a})$ phonon along ($H$1$\bar{1}$), $H\in$[4,5] for a subset of $\mathbf{Q}$ points.
\textbf{c} The $F_{4}(z,z)$ forces as fitted from the experimental data. The four data points are an average of fitting results to data sets taken under different ($H$1$\bar{1}$), $H\in$[3,4] and $H\in$[4,5] repsectively, but the same selected fluences.
The error bars reflect the standard deviation of these fitting results.
\textbf{d} The $F_{4}(x,x)$ forces as fitted from the experimental data.}    
	\label{fig:TA&LA fluence dependence}
\end{figure}  


The interatomic forces that determine materials structure and dynamics can be modified by temperature, pressure, chemical composition and applied fields leading to new equilibrium phases with dramatically different electronic, thermal and  mechanical properties ~\cite{born1955dynamical,brovnman1967phonon,sham1969electronic}. 
Methods such as inelastic X-ray\cite{ Krisch2007,Baron2014, Baron2014a} and neutron scattering\cite{Brockhouse1955, Cowley1964} are valuable because they can measure how the lattice dynamics change upon approaching or undergoing a phase transition.  
Ultrafast photo-excitation can also be used to generate non-equilibrium states on a time-scale that is short compared to various relaxation pathways. In some cases the excitation is only a small perturbation, and much can be learned about the equilibrium dynamics by studying how the material returns to equilibrium. In other cases, novel states of matter may be produced, with properties that do not exist in equilibrium~\cite{basov2017towards}. Either way,  conventional lattice dynamics measurements are not able to access the short time scales involved in the transition, and thus we need new tools for measuring the transient, excited state phonon dispersion.  
Time-domain, femtosecond X-ray diffuse scattering from free-electron lasers has recently been demonstrated to be able to give access to the phonon dispersion following ultrafast photoexciation, by measuring temporal coherences in the mean-square atomic displacements as a function of momentum transfer, associated with a rapid change in the interatomic forces\cite{Trigo2013a,zhu2015phonon,Jiang2016}.  In the case of photoexcited bismuth, Teitelbaum et al. \cite{Teitelbaum2021}, were able to measure changes in the interatomic forces associated with the well-established partial reversal of the Peierls distortion upon above gap excitation. Here we extend this technique to the study of a novel photo-induced lattice instability in the resonantly bonded monochalogenide, SnSe.  In this case, we have recently found through ultrafast X-ray diffraction, that SnSe undergoes a lattice instability towards a higher symmetry structure that is distinct from its well-known high temperature phase that is associated with the materials low thermal conductivity and high thermoelectric figure of merit\cite{Huang2022}. The new instability drives the material towards a structure that is an orthorhombic distortion of the cubic rock-salt structure which is important for its topological properties.  While time-resolved X-ray diffraction can yield changes in average structure of the material, it does not directly yield microscopic information on the changes in bonding, or the electronic states, that give rise to it.

At room temperature, the SnSe structure belongs to the orthorhombic space group Pnma. 
Its structure can be understood in terms of a distortion from the cubic parent (rocksalt) structure.  As shown in Fig.~\ref{fig:general_procedure} $\textbf{a}$, the six nearest neighbor bonds of the rocksalt structure yield four inequivalent bonds marked in different colors. 
In particular, the $d_2$ and $d_3$ bonds primarily along [011] and [01$\bar{1}$] directions are part of the resonant bonding network and lie approximately in the $\mathbf{b}$-$\mathbf{c}$ ($y$-$z$) plane, and originate largely from unsaturated $p$ orbitals. 
These resonant bonds have distinct characters from metallic, ionic or covalent bonds~\cite{Guarneri2021}. 
This resonantly bonded network exhibits long-range interatomic interactions,  large polarizability and large Born effective charges. Frozen phonon calculations have shown that in the equilibrium, soft phonon displacement induces long-range charge-density perturbations along resonant bonding directions~\cite{li2015orbitally}, which explains soft phonon behavior that leads to the thermal phase transition to a Cmcm structure (above 807~K~\cite{li2015orbitally,chattopadhyay1986neutron}), as well as large anharmonicity that leads to low thermoconductivity~\cite{Lee2014b}. The Pnma phase of SnSe features a stacking of bilayers along $\mathbf{a}$-axis. The distortion of its local structure as compared to an octahedral coordination in the rocksalt structure, is most obviously seen from the large tilting of $d_4$ bond that connects the bilayers. See Fig.~\ref{fig:general_procedure} $\textbf{a}$. 
The $d_4$ bond thus does not belong to a resonant bonding network.
By fitting an interatomic force model to the excited state phonon dispersions obtained in a time-resolved X-ray diffuse scattering measurement, we show that it is the change in interatomic interaction of the bilayer-connecting $d_4$ bond, rather than the in-plane resonant bonds (including $d_2$ and $d_3$), that destabilizes the photoexcited SnSe structure and leads to soft phonons.
These conclusions on interatomic bonding highlight the importance of nonequilibrium lattice dynamics measurements in addition to diffraction that only probes the average lattice structure.

Time-resolved diffuse scattering~\cite{Trigo2013a} requires a high X-ray photon flux of free electron lasers, as diffuse scattering efficiency is orders of magnitude lower than Bragg peaks.
We show the experimental setup in Fig.~\ref{fig:general_procedure}$\textbf{b}$, the pump pulse is centered at 1.55 eV, the probe beam is an X-ray pulse centered at 9.5 keV from the LCLS.
The X-ray beam impinges on the SnSe sample at grazing incidence to better match the pump and probe, given the large dispartity in their absorption.  
The crystal is oriented by rotating its azimuth angle $\varphi$, a two dimensional detector allows us to map out a large portion of the Ewald sphere.

In Fig.~\ref{fig:general_procedure} $\textbf{c}$ we show the time dependence of diffuse scattering intensity for selected scattering vectors, or momentum transfer, 
$\mathbf{Q}=(H1\bar{1})$ where $H\in[3,4]$ in reciprocal lattice units of the orthorhombic Pnma structure. The scattered X-ray intensity is modulated as a function of the pump-probe delay $t$. 
Fig.~\ref{fig:general_procedure} $\textbf{d}$ is a color plot showing the magnitude of the Fourier transform of time-domain data as shown in Fig.~\ref{fig:general_procedure} $\textbf{c}$, and clearly shows dispersive modes. 
We extract the frequencies of oscillations by a linear prediction (LP) method that decomposes the data into a sum of decaying cosines where the number of oscillators is determined directly from the data~\cite{barkhuijsen1985retrieval,epps2019singular1,epps2019singular2}.
The black lines in Fig.~\ref{fig:general_procedure} $\textbf{c}$ are sum of the LP components of the data, which correspond to known phonon modes of SnSe.

Fig.~\ref{fig:full_dispersion} combines data collected over the reciprocal space range $\mathbf{Q}=$(H$1\bar{1}$), ($H\in[3,5]$).  
The size of the dots represents the log-scaled amplitude of the oscillations from the LP, while the colors of the dots represent different optical pump fluences.
The solid lines 
show the phonon dispersion computed from density function theory (DFT) based calculations, which reproduce well those from inelastic neutron scattering~\cite{li2015orbitally,bansal2016phonon,Lanigan-Atkins2020}. 
The branches shown in red are the $\mathbf{c}$-polarized transverse acoustic (TA) branch which folds into the lowest transverse optical (TO) branch. From here on, these two branches are referred together as $T(\mathbf{c})$. Similarly, the blue line shows the $\mathbf{a}$-polarized longitudinal acoustic (LA) branch that folds into the lowest longitudinal optical (LO) branch (referred  together as $L(\mathbf{a})$).
The assignment of $L(\mathbf{a})$ and $T(\mathbf{c})$ phonons branches with reduced wavevectors $\mathbf{q}=(h00)$ (along $\mathbf{\Gamma} (h=0)$ $-$ $\mathbf{X} (h=0.5)$) is based on the polarization selectivity of the phonon structure factor.  We do not observe any oscillations in the diffuse scattering attributed to $\mathbf{b}$-polarized modes, and for clarity we do not include their calculated dispersion here.

We show in Fig.~\ref{fig:TA&LA fluence dependence}(a-b) the measured fluence dependence of the $T(\mathbf{c})$ and $L(\mathbf{a})$ mode frequencies for multiple $H$ values. 
The entire $T(\mathbf{c})$ branches soften with fluence, most significantly at zone center ($H$=4), and resembles the softening with temperature across the Pnma-Cmcm transition (see extended data figure). 
However, the frequency of $L(\mathbf{a})$ is non-monotonic with fluence, most pronounced near the avoided crossing seen in Fig. \ref{fig:full_dispersion}. The frequency of $L(\mathbf{a})$ hardens below around 1mJ/cm$^2$, and softens as the fluence keeps increasing.
In order to gain insight into which interatomic interactions are most responsible for the photoexcited lattice dynamics reflected in the changes of phonon frequencies, we fit an interatomic force model to the measured dispersion. 
We define the components of the pair-wise interatomic force tensor between two atoms connected by bond $d_n$, as $F_n(i,j)$, representing the force on one atom in the $i^\text{th}$ direction to a unit displacement of the other in the $j^\text{th}$ direction ($i,j \in{x,y,z}$).
We perform least-square fitting to the experimental $T(\mathbf{c})$ and $L(\mathbf{a})$ frequencies (Fig.~\ref{fig:full_dispersion}) by adjusting a subset of the $F_n(i,j)$.
The model includes $d_1$-$d_4$ (highlighted in Fig.~\ref{fig:general_procedure}$\textbf{a}$) 
as well as other longer range forces that are observed to have a relatively strong effect on the calculated $T(\mathbf{c})$ and $L(\mathbf{a})$ frequencies. Bonds not incorporated in the model remain fixed to the initial equilibrium values computed from DFT (see Method for details on the bond selection and fit procedure).

We find from our fitting results that a modification to a single force constant $F_4(x,x)$ dominates the observed changes in $L(\mathbf{a})$ frequencies and similarly $F_4(z,z)$ for $T(\mathbf{c})$.
Fig.~\ref{fig:TA&LA fluence dependence} $\textbf{c}$[$\textbf{d}$] shows the average value of $F_4(z,z)$ ($F_4(x,x)$) independently fit to the two data sets $H\in$ (3,4) and $H\in$ (4,5) in Fig.~\ref{fig:full_dispersion}.
Fig.~\ref{fig:TA&LA fluence dependence} $\textbf{c}$ suggests an increase of the initially negative $F_4(z,z)$ upon increased fluence, which is well correlated to the $T(\mathbf{c})$ softening under photoexcitation. 
The nonmonotonic fluence dependence of fitting results $F_{4}(x,x)$ as shown in Fig.~\ref{fig:TA&LA fluence dependence} $\textbf{d}$, is well correlated to the $L(\mathbf{a})$ fluence dependence behavior under photoexcitation (see Method for a statistical analysis of the fit results). 

In thermal equilibrium, the Pnma-Cmcm instability, and in turn, the softening of low-lying TO phonons, is governed primarily by the in-plane resonant bonds including $d_2$ and $d_3$, rather than the out of plane $d_4$~\cite{li2015orbitally,Lanigan-Atkins2020,Lee2014b,qin2016resonant}.
In fact, the most strongly affected Raman-active modes across the thermal Pnma-Cmcm transition are those polarized along $\mathbf{c}$ rather than $\mathbf{a}$~\cite{Lanigan-Atkins2020, Liu2018a}.
However, even though $d_2$, $d_3$ and other resonant bonds are included into the fitting model, 
the fitted force tensors for the resonant bonds do not correlate significantly with the mode softening. 
On the other hand, the fluence dependence of $d_4$ bond force constants has a good correlation with low frequency phonon. The weakening of inter-bilayer coupling suppresses the frequency of the $T(\mathbf{c})$ propagating along $(h00)$, and destabilizes the structure. 
The fitting results suggest that, under photoexcitation, resonant bonds are less responsible for phonon softening and lattice instability compared to the bilayer-connecting $d_4$ bond.

We infer that interatomic interactions between atoms connected by $d_4$ are modified by changes in occupation of the lone-pair orbital (the mixed orbital of cation $s$ and the out-of-plane chalcogen $p$), due to their large spatial overlap with $d_4$ bond~\cite{li2015orbitally}. 
For materials with rocksalt parent structure and resonant bonds, the lone pair orbital can have a large effect on structural distortions~\cite{rabe2007modern,waghmare2003first,orgel1959769}. 
Specifically for SnSe, such orbital breaks the parent cubic structure into bilayers and tilts $d_4$ bond away from $\bf{a}$-axis direction in Pnma. The lone pair orbital is located in the space opened up by the tilted bond, right next to $d_4$ bond~\cite{li2015orbitally}, see the black arrow in Fig.~\ref{fig:general_procedure}$\textbf{a}$. 
This is consistent with our previous diffraction studies where we find that under photoexcitation the bilayer-connecting $d_4$ bond tilts away from the high temperature Cmcm structure and towards a new Immm structure, and experiences the largest bond length change from the coherent atomic motion~\cite{Huang2022}. 
The time-resolved diffuse scattering measurements reported here thus suggest that the bond-level origin of the nonthermal lattice instability in SnSe is a direct consequence of the depopulation from lone pair orbitals.  
We note the first-principles, DFT calculations in ref ~\cite{Huang2022} also support the importance of de-occupation of lone-pair orbitals to the structural instability observed.
Finally, we attribute the unusual nonmonotonic fluence dependence of $F_{4}(x,x)$ as likely due to the details of the excitation density dependent spatial and temporal relaxation dynamics of the photoexcited carriers, which are known to suffer from  bottle-neck effects in other semiconductors\cite{Leheny1979,VanDriel1979,Yoffa1981}.


\begin{figure}
	\centering
	\includegraphics{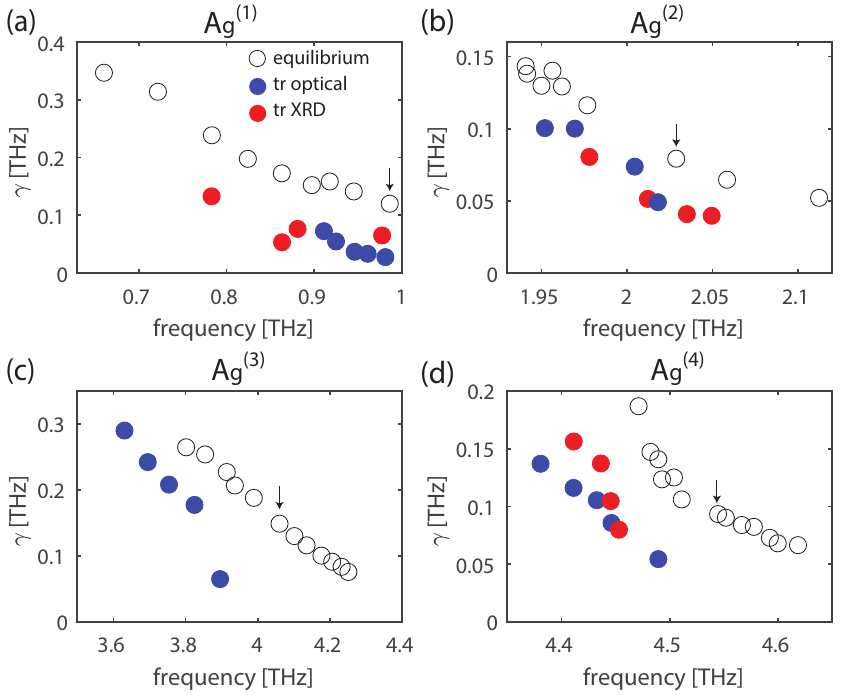}
	\caption{\textbf{Damping constants ($\gamma$) of the phonons compared between the photoexcited states and the thermal equilibrium phase.} $\gamma$ of photoexcited A$_g$ Raman modes [\textbf{a} (A$_g^{(1)}$), 
	\textbf{b} (A$_g^{(2)}$),
	\textbf{c} (A$_g^{(3)}$),
	\textbf{d} (A$_g^{(4)}$)] are obtained from either the inverse lifetime of time resolved X-ray scattering measurement (red dots), or time resolved optical reflectivity (blue dots). Phonon linewidths of photoexcited states are consistently lower  than the thermal equilibrium (black circle, measured by Raman scattering~\cite{Lanigan-Atkins2020,Liu2018a,Wu2020}) at the same given phonon frequencies. Black arrows correspond to the room temperature Raman measurements.
	}  
	\label{fig:linewidth}
\end{figure} 

Consistent with the nonthermal behavior, we show below that the photoexcited lattice is more harmonic than the thermal one at the temperature that produces similar level of softening of the symmetric Raman active $A_g$ modes at the zone center. 
We compare the damping versus the frequency change since the latter can be taken as a measure of the proximity to the phase transition~\cite{Dove1997}.
We define the damping rate $\gamma$ that describes the trajectory of a  damped harmonic oscillator (DHO) $x=Ae^{-\gamma t}\cos{(\omega_0 t+\phi)}$ in the time domain. 
The response function of the DHO measured in Raman spectroscopy is $\text{Im} \chi(\omega)=\frac{2\omega\gamma}{(\omega_0^2-\omega^2)^2+4\gamma^2\omega^2}$ on the frequency domain, for which the full width half maximum is $2\gamma$. 
In Fig.~\ref{fig:linewidth} (a-d) we show $\gamma$ for all four A$_g$ modes observed under photoexcitation (red dots are results from time-resolved XRD measurements and blue dots are from time-resolved optical reflectivity measurements) and measured by Raman spectroscopy in equilibrium (black open circles). 
For the same amount of softening, the phonon damping rate ($\gamma$) under photoexcitation are consistently lower than its counterpart under thermal equilibrium. 
The results are particularly notable given that the photoexcited carriers contribute to increasing the phonon linewidth due to the intraband phonon scattering~\cite{Fjeldly1973,Olego1981,Ledgerwood1996,Tsen2006}.
Thus the observations in Fig.~\ref{fig:linewidth} indicate that the photoexcited lattice is more harmonic than the lattice at a similar proximity (using the frequency softening as a metric) to the thermal phase Cmcm, which are consistent with our finding that photoinduced instability is not originated from resonant bonding network that gives rise to the large anharmonicity, but rather, the bilayer-connecting $d_4$ bond. 
Such observations have to do with the fact that  anharmonicity of a solid in thermal equilibrium are developed through a significant change in the lattice constants and internal atomic coordinates, which do not happen under photoexcitation on short time scales.

Clearly, X-ray scattering in the photoexcited states can establish the relation between the uniform ($\mathbf{q}=0$) structural distortions, the atomic bonding inferred from the ($\mathbf{q}\neq 0$)   transient excited state phonon dispersion, and the electron orbitals perturbed in the photoexcited states. 
Such relation can previously only be partially indicated in measurement of $\mathbf{q}=0$ macroscopic properties related to electron response function (e.g, Born effective charge, optical constants) in the photoexcited states, or the measurement of microscopic $\mathbf{q}\neq 0$ properties in the equilibrium states.
In SnSe, we identified that the photoexcited lattice instability is driven by changes in the bilayer-connecting bonds caused by nonequilibrium carrier distribution, instead of changes in resonant bonds which in thermal equilibrium produce soft and anharmonic phonons through long range charge perturbations.
The results have implications for ferroelectrics where the electron redistribution  significantly impacts the lattice dynamics, and phase change materials where the structural changes are coupled with bonding type change.
The methodology we demonstrate, can directly reveal the interplay between electron orbitals, atomic bonds and structural instability, and thus can be used to rationalize ways to control materials and design novel functional properties, for both equilibrium and nonequilibrium states.

\begin{acknowledgments}
Preliminary X-ray characterization was performed at beamline 7-2 at the Stanford Synchrotron Radiation Lightsource (SSRL). 
Y. H., S.T., G.d.P, D.A.R., A.M.L. and M.T. were supported by the U.S. Department of Energy, Office of Science, Office of Basic Energy Sciences through the Division of Materials Sciences and Engineering under Contract No. DE-AC02-76SF00515. S.Y. acknowledges support by the Fitzpatrick Institute for Photonics through a Chambers Scholarship. O.D. acknowledges support from the U.S. Department of Energy, Office of Science, Basic Energy Sciences, Materials Sciences and Engineering Division, under Award No. DE-SC0019978. Use of the LCLS and SSRL is supported by the US Department of Energy, Office of Science, Office of Basic Energy Sciences under Contract No. DE-AC02-76SF00515. Sample synthesis and characterization (A.F.M.) was supported by the U. S. Department of Energy, Office of Science, Basic Energy Sciences, Materials Sciences and Engineering Division. D.A.R. acknowledges discussions with Ivana Savic and Stephen Fahy.  
Y.H. acknowledges discussions with Xing He.
\end{acknowledgments}

\newpage

\section*{Methods}

\noindent \textit{Experimental Setup}

The experiment was performed at the X-ray pump probe (XPP) end-station at the Linac Coherent Light Source (LCLS) X-ray free-electron laser (FEL) using 9.5 keV X-rays\cite{chollet2015x,zhu2015phonon}. The X-rays were monochromatized (to 0.5~eV) using a diamond (111) double crystal monochromator, providing  nominally $> 10^9$ photons per pulse. The near infrared (NIR) pump pulses  with photon energy 1.55~eV were derived from a Ti:sapphire laser. The sample is a single crystal of SnSe grown with a Bridgman-type technique~\cite{li2015orbitally} and was polished with a [100] surface normal. The X-rays illuminated the sample at nominal grazing incidence of 0.5$^{\circ}$ with respect to the sample surface to approximatley match the penetration depth with the NIR laser, while the NIR beam was nearly colinear at an nominal incident angle of 1$^{\circ}$, with $p$ polarization. A fast scan delay stage controlled the nominal delay between the NIR and X-ray pulses. Scattered X rays were collected by the Cornell-SLAC pixel array detector (CSPAD)~\cite{hart2012cspad}. The total time resolution is $< 50$~fs, considering the duration of both pulses after correcting for their relative timing jitter. 
The relative arrival time $t$ between the X-ray probe and NIR pump was obtained on a shot-by-shot basis~\cite{harmand2013achieving}, and the X-ray scattering images were binned with intervals of 33~fs based on the sorted $t$. A two dimensional detector is placed $\sim 100$~mm behind the sample, which allows us to map out a large portion of the Ewald sphere. Different regions of the Brillouin zone were accessed by rotating the sample about the sample surface normal (azimuth $\varphi$), at nominally fixed grazing incidence.

\noindent \textit{Least Square Fit to the Photoexcited State Phonon Dispersion}

We assume that the interatomic interactions are renormalized by the photoexcited electron redistribution, without the change in lattice constants in the short probe time window, or significant changes in internal atomic coordinates, which is reported to be on the order of one thousandth of the lattice constant~\cite{Huang2022}. To fit the photoexcited phonon dispersions with interatomic forces, 
we allow a limited number of fit parameters to change while fixing others at values calculated from DFT.

We start from the interatomic force constant matrices in real space in the harmonic approximation.
Each $3\times 3$ matrix $F_n(i,j)$ can be diagonalized to obtain eigenvenctors eigenvalues. Note that the diagonalized real-space force matrix $F_n(i,j)$ is not a dynamical matrix. 
For $F_n(i,j)$ with real eigenvectors, one of the eigenvectors is approximately the vector connecting the two atoms, along the so-called bonding direction, while the other two are approximately perpendicular to the bonding directions . 
Since the atomic motiom is not large enough to change the symmetry of the crystal, or the symmetry of $F_n(i,j)$, we only vary the eigenvalues to minimize the least square errors in the fitting procedure, keeping the eigenvectors fixed.
The fitting parameters are the eigenvalue describing force along bonding direction ($f_n(\parallel )$) and the eigenvalues describing force perpendicular to bonding direction ($f_n(\perp)$).
We emphasize that $F_n$ represents for tensor while $f_n$ represents eigenvalues of the force tensor. 
For $F_4(i,j)$ that has imaginary eigenvalues (possibly due to the large effect of valence charge that destabilize the pair-wise ion-ion interaction), the corresponding eigenvectors cannot be related to bonding directions. 
The fit parameters of a $d_4$ bond are thus $F_4 (x,x)$, $F_4 (x,z)$, $F_4 (z,x)$, $F_4 (z,z)$. Based on the diffraction results under photoexcitation~\cite{Huang2022}, only atomic motion along $\textbf{a}$ and $\textbf{c}$ axis are observed. We thus assume that the change in force constants associated with atomic displacement along $\textbf{b}$-axis can be neglected.

\begin{figure*}
	\centering
	\includegraphics{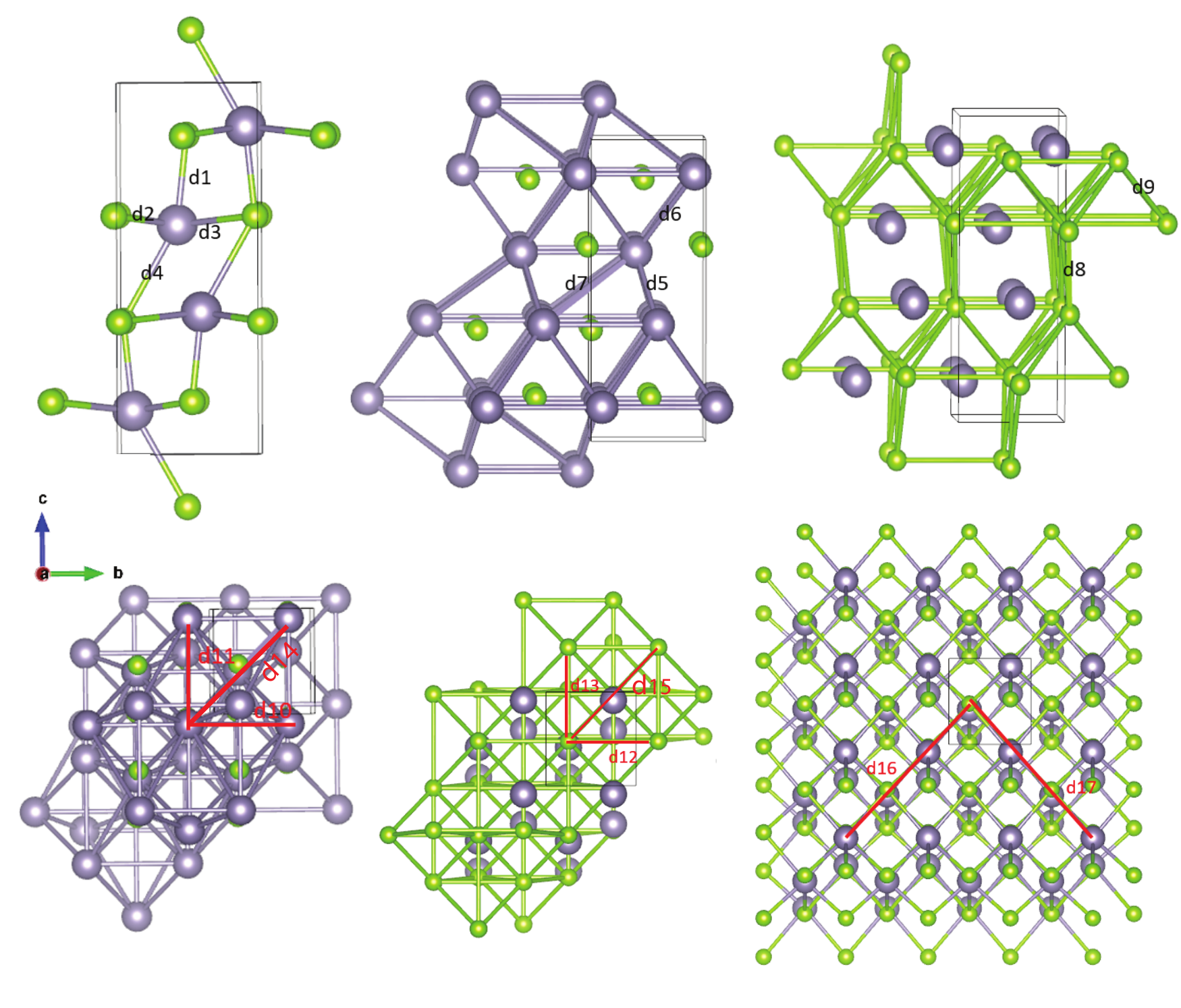}
	\caption{\textbf{Illustration of candidate bonds $d_1-d_{17}$.} }  
	\label{fig:candidate_bonds}
\end{figure*}

In Fig.S~\ref{fig:candidate_bonds} we display a total of 17 bonds. $d_{1}$ - $d_{13}$ are near neighbors chosen in the order of bond lengths. We also consider additional resonant bonds $d_{14}$ - $d_{17}$.
Apart from $d_{1}$ - $d_{4}$ that we will always include in the fitting model, we only select a subset of $F_n$,$(n=5-17)$, into the model. We include force constants into the fit in the following order of priority: $f_{1-3}$($\perp$ and $\parallel$) and $F_4 (i,j)$ ($i,j\neq y$), $f_7$($\perp$ and $\parallel$), $f_{11}$ ($\perp$ and $\parallel$), $f_{12}$ ($\perp$), $f_{10}$ ($\perp$), $f_{14}$ ($\perp$) and $f_{15}$ ($\parallel$). 
Bonds apart from $d_{1}$ - $d_{4}$ are sorted in the order of their covariance with the calculated phonon frequency. 
Including more fit parameters can only further drop the loss function (defined below) by less than 0.1\% and shows the sign of overfitting. 
Note that in the fit, the two $f_n(\perp)$ are constrained to change scale by a same scaling factor.

In order to perform the least square fit to the photoexcited phonon dispersion shown in Fig.~\ref{fig:full_dispersion}, we need to minimize the loss function, which is a sum of weighted squared error $\sum_{\textbf{q}}W_{{\textbf{q}}}(\omega^{calc}_{\textbf{q}}-\omega^{exp}_{\textbf{q}})^2$. 
The weight factor $W_{{\textbf{q}}}$ of each data point ($\textbf{q},\omega^{exp}_{\textbf{q}}$) is the signal noise ratio of the time domain trace sampled at $\textbf{q}$ (the noise level of the data can be extracted using the well defined algorithm developed in ref~\cite{epps2019singular1,epps2019singular2}). 

\noindent\textit{Systematic Analysis of Fit Results}

\begin{figure*}
	\centering
	\includegraphics{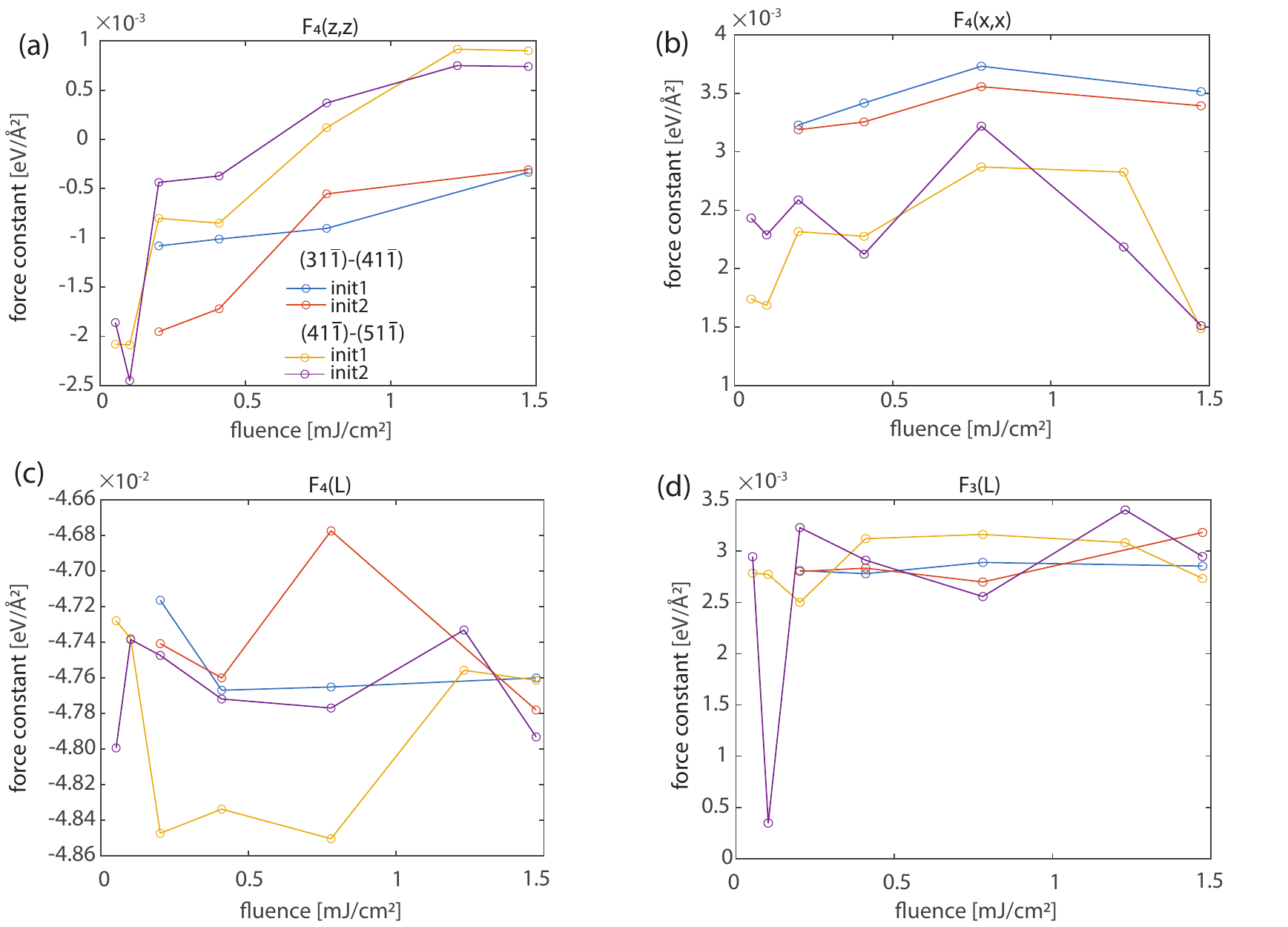}
	\caption{\textbf{The fitted force constants as a function of fluence.} For each force constant, the fit is performed given different reciprocial space regions and different initial fit parameters (init1 and init2).
	}  
	\label{fig:fitforce}
\end{figure*}

\begin{figure*}
	\centering
	\includegraphics{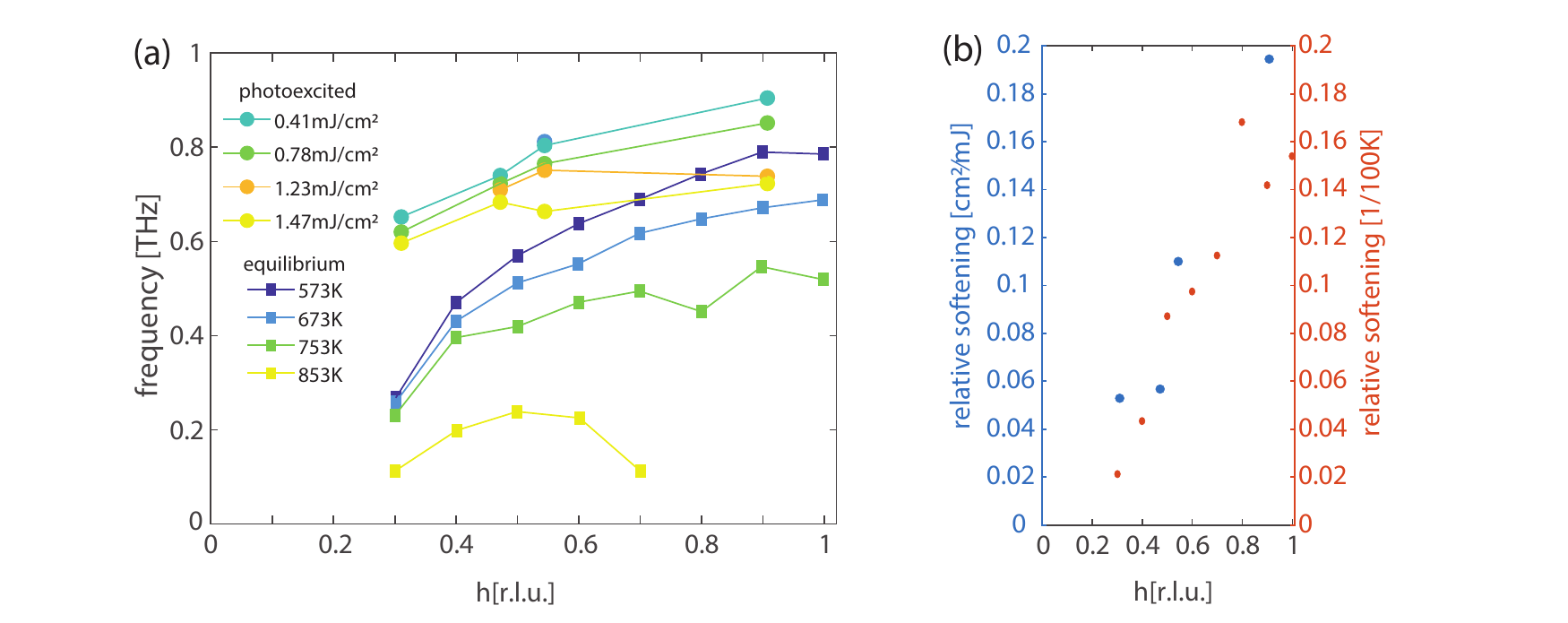}
	\caption{
	\textbf{$T(\mathbf{c})$ frequency dependence on fluence and temperature.}
	$\textbf{a}$ Photoexcited and thermal equilibrium $T(\mathbf{c})$ phonon dispersions. 
	$\textbf{b}$ The relative softening of $T(\mathbf{c})$ phonon, shown for both photoexctiation (defined as relative frequency change per mJ/cm$^2$), thermal equilibrium (defined as relative frequency change per Kelvin). Equilibrium data are taken from ref~\cite{Chatterji2018}.
	}   
	\label{fig:TA_addtional}
\end{figure*}

It is expected that a robust correlation between the fluence-dependent fitted force constants and the fluence-dependent phonon frequencies is manifested in various fit settings. 
In Fig.S~\ref{fig:fitforce} we show the fit results for four force constants, $F_4(x,x)$, $F_4(z,z)$, $F_2(\parallel)$, $F_3(\parallel)$, under a series of fluences. Apparenetly, $F_4(z,z)$ shows the same monotonic trend and $F_4(x,x)$ shows the same nonmonotonic trend for different fit settings, and the trends are similar to the fluence dependence of $L(\mathbf{a})$ and $T(\mathbf{c})$ respectively. 
Below we evaluate the correlation between fitted force constants  and observed phonon frequencies. 
We define target function $Y$ (for either $L(\mathbf{a})$ or $T(\mathbf{c})$) that considers the phonon frequencies of $\textbf{q}$ points across $\Gamma$-$X$-$\Gamma$ instead of individual $\textbf{q}$ points. 
The target function for $T(\mathbf{c})$ is the sum of product of frequency change and the first reduced Miller index $h$ value of $\textbf{q}$ points $Y=\sum_{\textbf{q}}\Delta\omega_{\textbf{q}}\times h_{\textbf{q}}$, based on the linear relationship between $\Delta \omega_{\textbf{q}}$ and $h_{\textbf{q}}$. See Fig.S~\ref{fig:TA_addtional}.
The target function for $L(\mathbf{a})$ is defined as the sum of absolute change of $L(\mathbf{a})$ frequency across $\Gamma$-$X$-$\Gamma$, $Y=\sum_{\textbf{q}}\Delta\omega_{\textbf{q}} $, based on the non-monotonic fluence dependence of $L(\mathbf{a})$ across $\Gamma$-$X$-$\Gamma$.
We will compare, under different settings, $r_i$, the normalized correlation between X$_i$ (fitted force constant, one of $f_{1-3}$($\perp$ and $\parallel$) and $F_4 (i,j)$ ($i,j\neq y$), $f_7$($\perp$ and $\parallel$), $f_{11}$ ($\perp$ and $\parallel$), $f_{12}$ ($\perp$), $f_{10}$ ($\perp$), $f_{14}$ ($\perp$) and $f_{15}$ ($\parallel$)) and Y (target function related to the observed phonon frequency). 
When $r_i$ is near 1 or $-1$, the linear correlation is strong, when it is near 0, the linear correlation is weak.
By definition, positive $r_i$ means a positive correlation, and negative $r_i$ a negative correlation. 
We also compare the linear coefficient $k_i$ assuming linear relation between Y and X$_i$, under different fit settings. 
A robust correlation between X$_i$ and Y would imply that under various settings (different initial fit parameters in the least square fitting, different data sets taken with different Brillioun zone cuts, different selections of fit parameters that contain X$_i$) the $|r_i|$ is consistently large,  $r_i$ in all these cases have the same sign, and that $k_i r_i$ is large. 

\begin{figure*}
	\centering
	\includegraphics{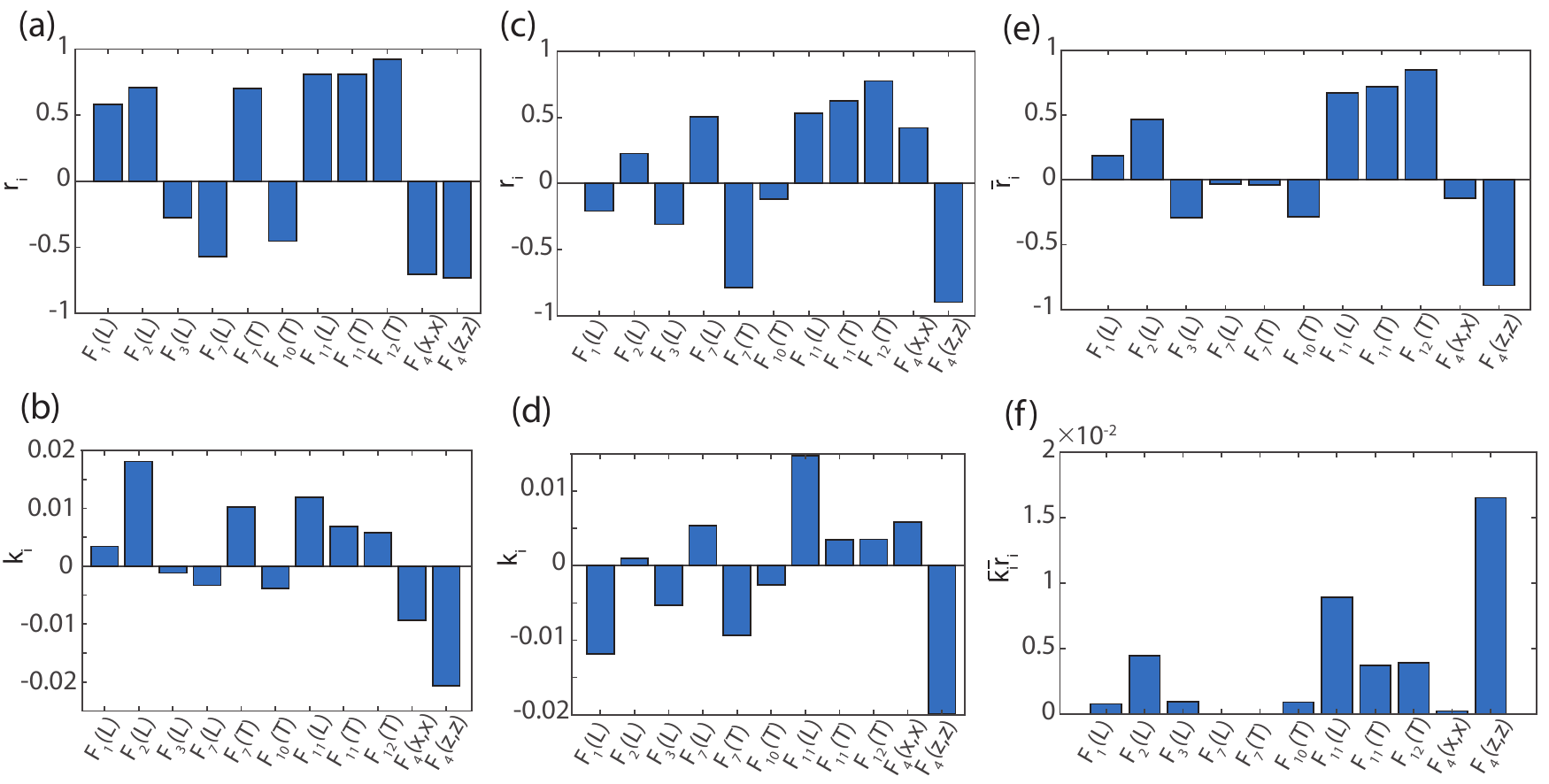}
	\caption{\textbf{Systematic analysis of $T(\mathbf{c})$ fits to measurements on different Brillouin zones.} 
	$\textbf{a}$ $r_i$ for the fit to data set taken along (31$\bar{1}$)-(41$\bar{1}$). $r_i$ is the correlation between Y and X$_i$. 
	$\textbf{b}$ $k_i$ for the fit to data set taken along (31$\bar{1}$)-(41$\bar{1}$). $k_i$ is the linear slope assuming a linear function of Y versus X$_i$. 
	$\textbf{c}$ $r_i$ for the fit to data set taken along (41$\bar{1}$)-(51$\bar{1}$).
	$\textbf{d}$ $k_i$ for the fit to data set taken along (31$\bar{1}$)-(41$\bar{1}$).
	(e) Average of $\textbf{a}$ and $\textbf{c}$, denoted as $\bar{r}_i$. 
	(f) Average $\bar{k}_i$ of $\textbf{b}$ and $\textbf{d}$, multiplied by $\bar{r}_i$.
	}  
	\label{fig:systematics_TA_geometry}
\end{figure*}

\begin{figure*}
	\centering
	\includegraphics{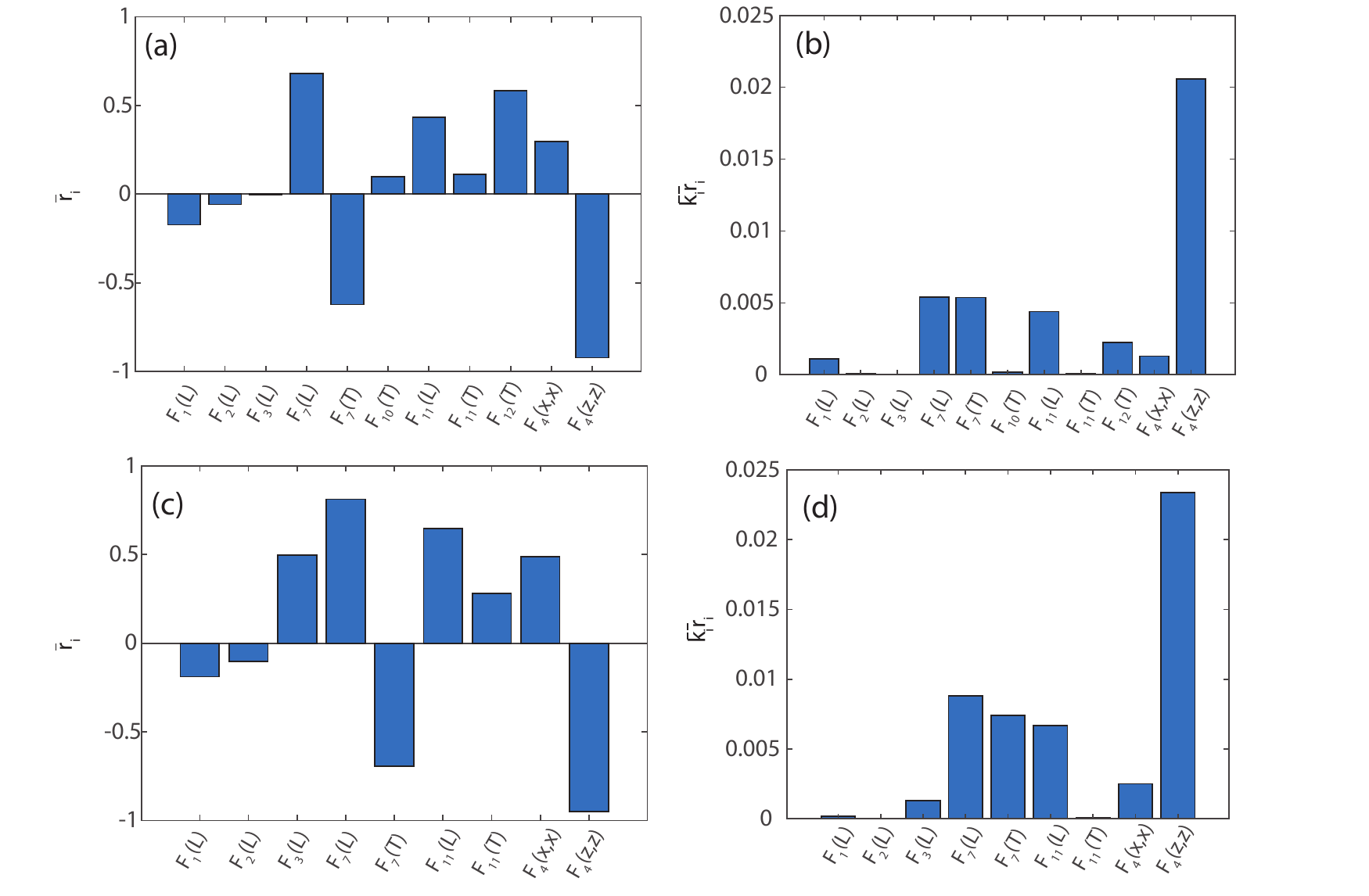}
	\caption{\textbf{Systematic analysis of $T(\mathbf{c})$ fits given different initial fit parameters, and using different force constant subsets.}
	$\textbf{a}$ $\bar{r}_i$ averaging ${r}_i$ of two fits using different initial fit parameters. 
	$\textbf{b}$ $\bar{k}_i\bar{r}_i$ multiplying the average quantities $\bar{k}_i$ and $\bar{r}_i$ between two fits using different initial fit parameters. 
	$\textbf{c}$ $\bar{r}_i$ averaging ${r}_i$ of two fits that both include fit parameters as shown on the $x$-axis.
	$\textbf{b}$ $\bar{k}_i\bar{r}_i$ multiplying the average quantities $\bar{k}_i$ and $\bar{r}_i$ between two fits that both include fit parameters as shown on the $x$-axis.
	}  
	\label{fig:systematics_TA_init_config}
\end{figure*}

We perform the systematic analysis of $T(\mathbf{c})$ target function (Y) and bond parameters (X$_i$), for data taken on different part of the reciprocal space, see Fig.S~\ref{fig:systematics_TA_geometry}. 
We further show the systematic analysis of $T(\mathbf{c})$ given different initial least square fit parameters, as well as different fit parameter selections, see Fig.S~\ref{fig:systematics_TA_init_config}.
From these plots, we conclude that $F_4(z,z)$ best explains the $T(\mathbf{c})$ frequency, because both the quantities $|r_i|$ and $k_i r_i$ are consistently large.

\begin{figure*}
	\centering
	\includegraphics{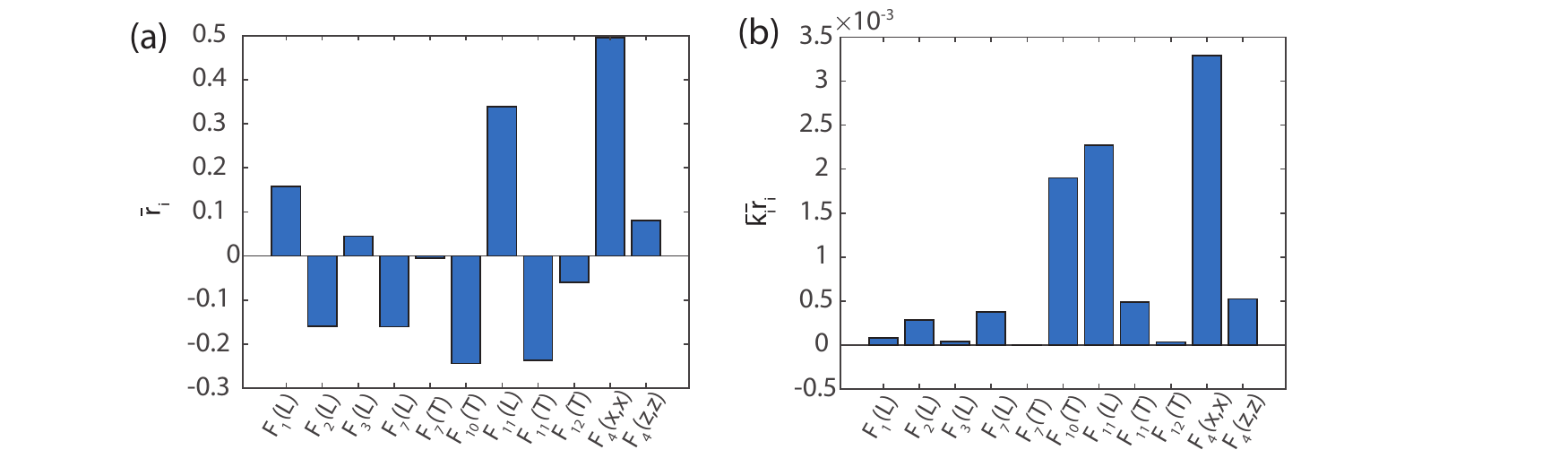}
	\caption{\textbf{Systematic analysis of $L(\mathbf{a})$ fit results}. 
	$\textbf{a}$ $\bar{r}_i$ the average of $r_i$ of four fits under two initial conditions of the fit, and two different parts of the reciprocal space, $\mathbf{Q}$ path (31$\bar{1}$)-(41$\bar{1}$) and (41$\bar{1}$)-(51$\bar{1}$).
	$\textbf{b}$ $\bar{k}_i \bar{r}_i$ where both $\bar{k}_i $ and $\bar{r}_i$ are over the four fits.
	}  
	\label{fig:systematics_LA}
\end{figure*}

In Fig.S~\ref{fig:systematics_LA} we do a similar analysis for $L(\mathbf{a})$ phonon and conclude the $F_4(x,x)$ best explains the $L(\mathbf{a})$ frequency.

\noindent\textit{Density Functional Theory}

DFT  was performed using VASP, with the projected-augmented-wave (PAW) and local density approximation (LDA)~\cite{kresse1996efficient, blochl1994projector, kresse1999ultrasoft}, which proves to yield accurate phonon dispersions~\cite{li2015orbitally, bansal2016phonon, Lanigan-Atkins2020} and provides better agreement with INS and Raman measurements than the Perdew-Burke-Ernzerhof (PBE) generalized gradient approximation (GGA)~\cite{bansal2016phonon, Lanigan-Atkins2020}. The Pnma equilibrium structure is relaxed with a kinetic energy cutoff of 500~eV and an electronic Monkhorst-Pack grid with $6 \times 12 \times 12$ $k$-points, giving lattice constants ($a=11.31$\,\AA\, $b= 4.12$\,\AA\, $c=4.30$\,\AA\,) in good agreement with X-ray diffraction experimental report at 296~K: \cite{wiedemeier1979thermal} ($a=11.50$\,\AA\, $b=4.16$\,\AA\, $c=4.45$\,\AA\,) and \cite{sist2016crystal} ($a=11.44$\,\AA\, $b=4.13$\,\AA\, $c=4.45$\,\AA\,).

\textbf{Data availability} The data that support the findings of this study is available from the corresponding author upon reasonable request.

\textbf{Supplementary Information} Supplementary information is available in the online version of the paper.

\textbf{Author Contributions} Y.H., S.T., G.D.P., T.S., M.C., D.Z., J.N.L., D.B., O.D., D.R., M.T. participated in the experiment, Y.H. performed the data analysis and wrote the paper with input from all the authors, S.T., D.R, M.T. gave suggestions on the data analysis, S.Y. performed DFT calculations.

\textbf{Author Information} The authors declare no competing financial interests.


\bibliography{aSnSe_DS}
\newpage
\newpage

\newpage


\newpage

\newpage

\newpage

\end{document}